\documentclass{PoS}

\title{Global Fit of $\alpha_s(m_Z)$ to Thrust at N${}^3$LL Order with Power Corrections}

\ShortTitle{Global Fit of $\alpha_s(m_Z)$ to Thrust at NNNLL Order with Power Corrections}

\author{Riccardo Abbate\\
Center for Theoretical Physics, Massachusetts Institute of
Technology, Cambridge, MA 02139\\
 E-mail: \email{rabbate@mit.edu}}

\author{Michael Fickinger\\
Department of Physics, University of Arizona, Tucson, AZ 85721\\
 E-mail: \email{fickinger@physics.arizona.edu}}

\author{\speaker{Andre Hoang}
\\
Max-Planck-Institut f\"ur Physik (Werner-Heisenberg-Institut),
F\"ohringer Ring 6, 80805 Munich\\
E-mail: \email{ahoang@mppmu.mpg.de}}

\author{Vicent Mateu\\
Max-Planck-Institut f\"ur Physik (Werner-Heisenberg-Institut),
F\"ohringer Ring 6, 80805 Munich\\
E-mail: \email{mateu@mppmu.mpg.de}}

\author{Iain W.\ Stewart\\
Center for Theoretical Physics, Massachusetts Institute of Technology, 
Cambridge, MA 02139\\
E-mail: \email{iains@mit.edu}}

\abstract{ From soft-collinear effective theory one can derive a factorization
  formula for the $e^+e^-$ thrust distribution $d\sigma/d\tau$ with $\tau=1-T$
  that is applicable for all $\tau$.  The formula accommodates available ${\cal
    O}(\alpha_s^3)$ fixed-order QCD results, resummation of logarithms at
  N${}^3$LL order, a universal nonperturbative soft function for hadronization
  effects, factorization of nonperturbative effects in subleading power
  contributions, bottom mass effects and QED corrections. We emphasize that the
  use of Monte Carlos to estimate hadronization effects is not compatible with
  high-precision, high-order analyses.  We present a global analysis of all
  available $e^+e^-$ thrust data measured at $Q=35$ to $207$~GeV in the tail
  region, where a two-parameter fit can be carried out for $\alpha_s(m_Z)$ and
  $\Omega_1$, the first moment of the soft function. To obtain small theoretical
  errors it is essential to define $\Omega_1$ in a short-distance scheme, free
  of an ${\cal O}(\Lambda_{\rm QCD})$ renormalon ambiguity.  We find
  $\alpha_s(m_Z)=0.1135 \pm (0.0002)_{\rm expt} \pm (0.0005)_{\Omega_1} \pm
  (0.0009)_{\rm pert}$ with $\chi^2/{\rm dof}=0.9$.  }

\FullConference{RADCOR 2009 - 9th International Symposium on Radiative Corrections (Applications of Quantum Field Theory to Phenomenology),
                 October 25-30 2009,
                 Ascona, Switzerland\\
Preprint: MIT-CTP 4118, MPP-2010-8}

\begin{document}

A traditional method for testing the theory of strong interactions (QCD) and 
to make precise determinations of the strong coupling $\alpha_{s}$
is the analysis of event-shapes measured at $e^{+}\, e^{-}$ colliders~\cite{Kluth:2006bw}. 
One of the most frequently studied event-shape variables is thrust~\cite{Farhi:1977sg}
\begin{equation}
\label{Tdef}
T \, = \,\mbox{max}_{\hat {\bf t}}\ \frac{\sum_i|\hat {\bf t}\cdot\vec{p}_i|}
{\sum_i|\vec{p}_i|}
\,,
\end{equation}
where the sum $i$ is over all final-state hadrons with momenta $\vec{p}_i$, and
the unit vector ${\hat {\bf t}}$ that maximizes the RHS of Eq.~(\ref{Tdef})
defines the thrust axis.  It is convenient to use the variable $\tau = 1-T$.
For the production of a pair of massless quarks at tree level $d\sigma/d\tau
\propto \delta(\tau)$, so the measured distribution for $\tau>0$ involves gluon
radiation and is highly sensitive to the value of $\alpha_s$. For $\tau$ values
close to zero the event has two narrow pencil-like, back-to-back jets, carrying
about half the center-of-mass (c.m.) energy into each of the two hemispheres
defined by the plane orthogonal to ${\hat {\bf t}}$. For $\tau$ close to the
kinematic endpoint $0.5$, the event has an isotropic multi-particle final state
containing a large number of low-energy jets.
The 
thrust distribution can be divided into three regions,
\begin{eqnarray}
   \mbox{\rm peak region:} &  \tau  \sim 2\Lambda_{\rm QCD}/Q \,, \nonumber\\
   \mbox{tail region:} & \quad 2\Lambda_{\rm QCD}/Q  \ll \tau < 1/3 \,, \nonumber\\
   \mbox{far-tail region:} &  1/3 \lesssim \tau \le 1/2 \,. \nonumber
\end{eqnarray}
For $\tau< 1/3$ the dynamics is governed by three
different scales. The \emph{hard scale} $\mu_H\simeq Q$, set by the $e^+e^-$
c.m.~energy $Q$, the \emph{jet scale}, $\mu_J\simeq Q\sqrt{\tau}$, the
typical momentum transverse to ${\hat {\bf t}}$ of the particles within each of
the two hemispheres, and the  \emph{soft scale} $\mu_S\simeq Q\,\tau$, the typical
energy of soft radiation between the hard jets.
In the {\em peak region} the distribution shows a strongly peaked maximum.
Since $\tau\ll 1$ one needs to sum 
large (double) logarithms, $(\alpha_s^j\ln^k\!\tau)/\tau$, and 
$d\sigma/d\tau$ is affected at leading order by a nonperturbative distribution, called
soft function $S_\tau^{\rm mod}$.  
In the analysis presented in this talk we consider the \emph{tail region}.  It
is populated predominantly by broader dijets and 3-jet events.  Here the three
scales are still well separated and one still needs to sum logarithms, but now
$\mu_S\gg \Lambda_{\mathrm{QCD}}$ so soft radiation can be described by
perturbation theory and the first moment of the soft function 
$\Omega_1=\int{\rm d}k (k/2)S_\tau^{\rm mod}(k-2\,\bar\Delta)$. 
Many previous event-shape analyses have relied on Monte-Carlo (MC) generators to quantify the
size of nonperturbative corrections. This is problematic since the partonic
contributions implemented in MC generators are (i) based on LL parton showers
and (ii) contain an infrared cut below which the perturbative parton shower is
switched off and replaced by hadronization models that are not derived from
QCD. Thus MC hadronization effects are not compatible with high-order
perturbative event-shape predictions using the common $\overline{\mbox{MS}}$
scheme.

In this talk we present a new analysis of $e^+e^-$ thrust data using the
soft-collinear effective theory (SCET), an effective theory for
jets~\cite{Bauer:2002nz}, to derive the theoretical QCD prediction of the thrust
distribution. Within SCET it is possible to formulate a factorization theorem
that allows to describe the thrust distribution for all $\tau$. The formula we
use is~\cite{AFHMS1}:
  \begin{equation} \label{masterformula}
\frac{\rm{d}\sigma}{\rm{d}\tau}  
= \! \int\!\! \rm{d}k\left(
\frac{\rm{d}\hat{\sigma}_{\rm s}}{\rm{d}\tau}+
\frac{\rm{d}\hat{\sigma}_{\rm ns}}{\rm{d}\tau}+
\frac{\Delta\rm{d} \hat{\sigma}_{b}}{\rm{d}\tau}\!\right)
\!\!\bigg(\!\tau- \frac{k}{Q}
\bigg) 
S_{\tau}^{\rm{mod}}(k\!-\!2\,\bar\Delta)
\times 
 \Big[ 1 + {\cal O}\Big(\alpha_s \frac{\Lambda_{\rm QCD}}{Q}\Big)\Big]
 \,.
\end{equation} 
\begin{figure*}[t!]
\includegraphics[scale=0.85]{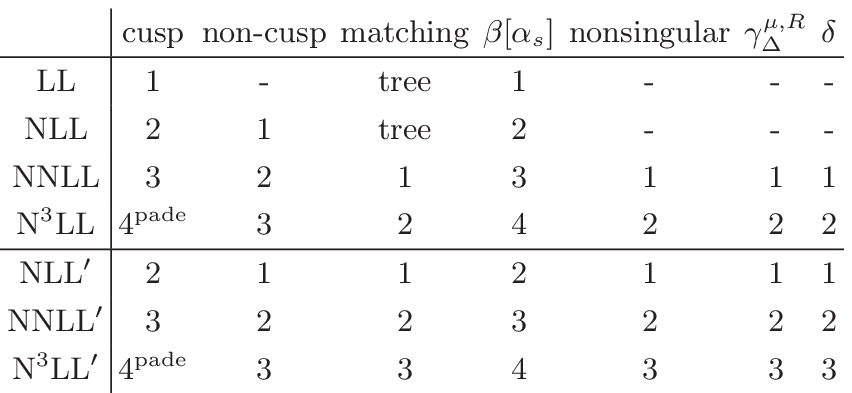}\hfill
\includegraphics[scale=0.85]{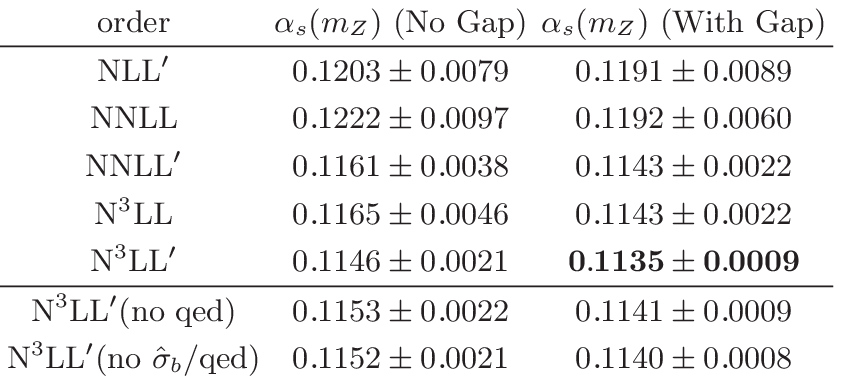}
\caption{(a) Ingredients for primed and unprimed orders used in our
  analysis. The numbers give the loop orders for the cusp and non-cusp anomalous
  dimensions, matching/matrix element contributions, the $\alpha_s$-running,
  the nonsingular distribution, the gap-anomalous dimensions, and the
  perturbative $R$-scheme
  subtractions $\delta$ for our scheme for $\Omega_1$. The 4-loop cusp anomalous
  dimension required at N${}^3$LL${}^\prime$ order is estimated from Pad\'e
  approximants. The associated uncertainty is negligible.
  (b) Central values and theory uncertainties for
  the fits at the different orders with and without the gap and renormalon
  subtractions.
  \label{fig:orderserrors}}
\end{figure*}
Due to lack of space we describe in the following only the main features of
Eq.~(\ref{masterformula}).  For details, explicit analytic expressions, how our
implementation improves upon earlier analyses in the literature, and a complete set
of references we refer the reader to Ref.~\cite{AFHMS1}.  The term
$\mathrm{d}\hat{\sigma}_{\rm s}/\mathrm{d}\tau$ contains the {\bf singular
  partonic contributions}. It factorizes further into a hard coefficient, a jet
function and a partonic soft function governed by the renormalization scales
$\mu_{H}$, $\mu_{J}$ and $\mu_{S}$, respectively, and renormalization group (RG)
evolution factors that sum logarithms between the hard, jet and soft scales.
Using results from the existing literature, SCET allows to sum the logarithms at
N${}^3$LL order~\cite{Becher:2008cf}, which is two orders beyond the classic
resummation method~\cite{Catani:1992ua} that is valid up to NLL order.  The jet
and partonic soft functions contain $\alpha_s^j\,[\ln^k(\tau)/\tau]_+$ and
$\alpha_s^j\,\delta(\tau)$ distribution terms. They are known to ${\cal
  O}(\alpha_s^2)$, and at ${\cal O}(\alpha_s^3)$ all logarithmic terms are known
from the renormalization group. Two unknown ${\cal O}(\alpha_s^3)$
non-logarithmic constants contribute to the theory error in our highest order
numerical analysis.  The hard function in our analysis is fully known at ${\cal
  O}(\alpha_s^3)$~\cite{Baikov:2009bg,Lee:2010cg} and also includes the axial-vector
singlet contributions at ${\cal O}(\alpha_s^2)$.  To achieve a definition of the
soft function moment $\Omega_1$ that is free of a $\Lambda_{\rm QCD}$ renormalon
ambiguity, $\mathrm{d}\hat{\sigma}_{\rm s}/\mathrm{d}\tau$ contains subtractions
that eliminate partonic low-momentum
contributions~\cite{Hoang:2007vb,Hoang:2008fs}.  This requires the introduction
of the additional scale-dependent model parameter $\bar\Delta(\mu_R)$ 
(with $\mu_R\sim\mu_S$), called
the {\bf gap parameter}, visible in Eq.~(\ref{masterformula}).  In our numerical
tail-data fits $\bar\Delta(\mu_R)$ is contained in $\Omega_1$. The evolution of
$\bar\Delta(\mu_R)$ follows a new type of infrared RG equation formulated in
Refs.~\cite{Hoang:2008yj}.  We have also included final-state QED matrix
elements and QED RG corrections at NNLL order, derived from the QCD results.
The term $\mathrm{d}\hat{\sigma}_{\rm ns}/\mathrm{d}\tau$, called the {\bf
  nonsingular partonic distribution}, contains the thrust distribution in
strict fixed-order expansion up to ${\cal O}(\alpha_s^3)$ with the singular
terms contained in $\mathrm{d}\hat{\sigma}_{\rm s}/\mathrm{d}\tau$ subtracted to
avoid double counting. At ${\cal O}(\alpha_s)$ the nonsingular distribution is
known analytically, and at ${\cal O}(\alpha_s^2)$ and ${\cal O}(\alpha_s^3)$ we
rely on numerical results obtained from the programs EVENT2~\cite{Catani:1996jh}
and EERAD3~\cite{GehrmannDe Ridder:2007bj} (see also~\cite{Weinzierl:2009yz}). To achieve a consistent behavior in
the far-tail region infrared subtractions need to be implemented here as well.
A list of the perturbative ingredients for the different orders we consider is
given in Tab.~\ref{fig:orderserrors}a. N${}^3$LL${}^\prime$ is the highest order
we consider and contains all currently available perturbative information.
Finally, $\Delta {\rm d}\hat\sigma_b/{\rm d}\tau$ contains corrections to the
singular and nonsingular distributions due to the {\bf finite $b$-quark mass},
using Refs.~\cite{Fleming:2007qr} for the consistent treatment and resummation for the singular
terms. The entire partonic distribution is convoluted with the {\bf soft
  function} $S_{\tau}^{\rm mod}$ that describes the nonperturbative effects
coming from large-angle soft radiation and can be determined from experimental
data.  The last term in the brackets indicates the parametric size of the
dominant power corrections not contained in the factorization formula. For a
proper summation of large logarithmic terms it is necessary to adopt
$\tau$-dependent {\bf profile functions} for the renormalizations scales
$\mu_{H}$, $\mu_{J}$, $\mu_{S}$ and $\mu_R$ that follow the scaling arguments
given above.  For $\tau\to 0.5$ all profile functions need to merge into the
hard scale $\mu_H$ to ensure that in the large-$\tau$ endpoint region the
partonic distribution coincides with the fixed-order result, so that it does not
violate the proper behavior at multi-jet thresholds.  The variations of these
profile functions estimate higher order perturbative uncertainty, and
constitute our major source of theory uncertainty.

\begin{figure*}[t!]
\includegraphics[scale=0.85]{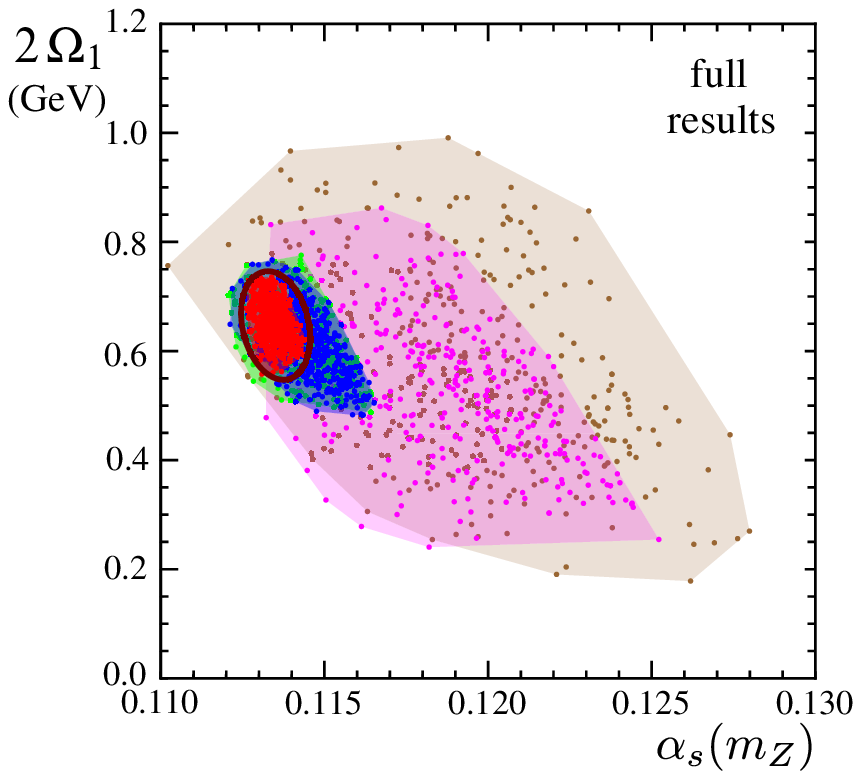}
\includegraphics[scale=0.85]{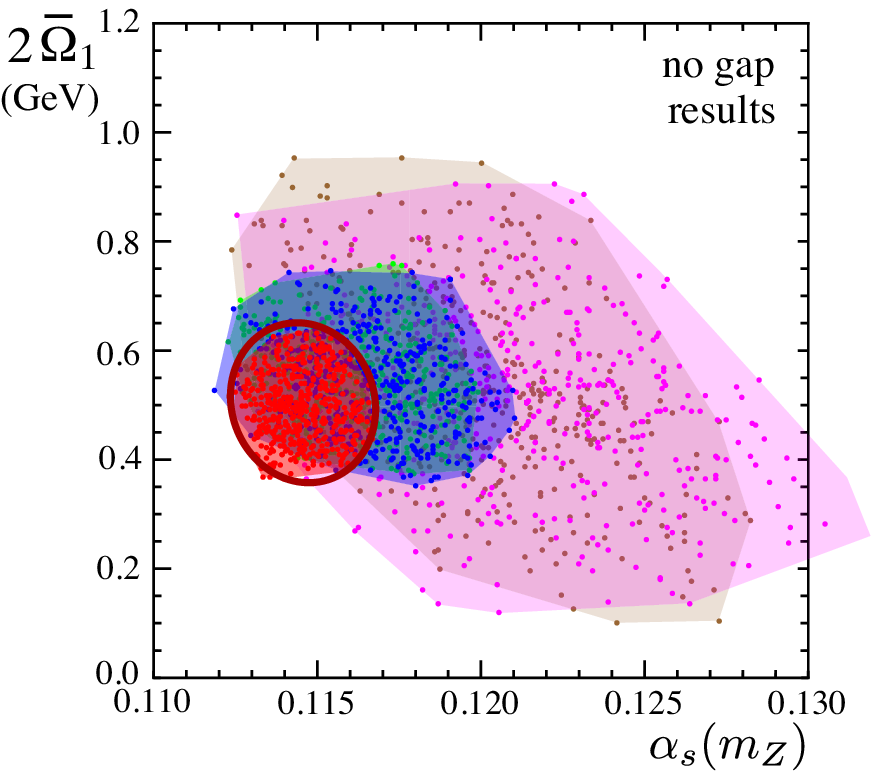}
\caption{Plots of $\Omega_1$ vs $\alpha_s(m_Z)$. (a) Includes perturbation
  theory, resummation of the logs, the soft model function and $\Omega_1$ with
  renormalon subtractions at $\mu_R=2$~GeV.  
  (b) As (a) but in a scheme $\bar\Omega_1$ without
  a gap, which gives perturbative results without the corresponding renormalon
  subtractions. 
  The shaded regions indicate the theory errors
  at NLL${}^\prime$ (brown), NNLL (magenta), NNLL${}^\prime$ (green), 
  N${}^3$LL (blue), N${}^3$LL${}^\prime$ (red). The dark red ellipses in (a) and (b)
  represent the $(\chi^2_{\rm min}+1$) error ellipses for the combined
  theoretical, experimental and hadronization uncertainties. The ellipse in (a)
  is displayed again in Fig.~3b.
  The  best fit points at N${}^3$LL${}^\prime$ with gap and renormalon
  subtractions shown in red in (a) each have $\chi^2/{\rm dof}\simeq 0.90$.
  \label{fig:M1alphagap}}
\end{figure*}

In our analysis we fit the factorization formula~(\ref{masterformula}) in the
tail region to all available $e^+e^-$ {\bf thrust data} from c.m.\ energies $Q$
between $35$ and $207$~GeV. In the tail region the distribution can be expanded
in $\Lambda_{\rm QCD}/(Q\tau)$ and thus described to high precision using
$\alpha_s(m_Z)$ and $\Omega_1$. We carry out a two-parameter fit for these two
variables. Fitting for $\Omega_1$ accounts for hadronization effects in a
model-independent way.  For the {\bf fitting procedure} we use a
$\chi^2$-analysis, where we combine the statistical and the systematical
experimental errors into the correlation matrix, treating the statistical errors
as independent. We also account for experimental correlations of thrust bins
obtained at one $Q$ value by one experiment through the minimal overlap model,
and find a similar central value to a completely uncorrelated treatment.
To estimate the {\bf theoretical errors} in the $\alpha_s-\Omega_1$ plane we
carry out independent fits for $500$ different sets of theory parameters (for
two unknown ${\cal O}(\alpha_s^3)$ non-logarithmic constants, the four-loop cusp
anomalous dimension, numerical uncertainties for the ${\cal
  O}(\alpha_s^{2,3})$ nonsingular distributions, parameters of the profile
functions/renormalization scales) which are randomly chosen in their natural
ranges with a flat distribution.  We take the area covered by the points of the
best fits in the $\alpha_s-\Omega_1$ plane as the theory uncertainty.

The result of our fits for our default thrust tail range $6/Q\le\tau\le 0.33$
(487 bins), at the five different orders we consider is displayed in
Fig.~\ref{fig:M1alphagap}. The left panel shows the results including the gap
and renormalon subtractions and the right panel without the gap and renormalon
subtractions. Each dot corresponds to a best fit for a given set of theory
parameters. The shaded areas envelop the best fit points and give the theory
uncertainties. The numbers for central values and theory errors at each order
are collected in Tab.~\ref{fig:orderserrors}b and also display the size of the
QED and $b$-quark mass effects.  We see the excellent convergence of the fit
results and the decrease of the respective theory uncertainties with increasing
perturbative order. Moreover, including the gap and the renormalon subtractions
leads to uncertainties that are about a factor of two smaller at the highest
three orders. This illustrates the impact of the renormalon contributions and
the necessity to subtract them from the partonic distribution. Our scan method
is more conservative than the traditional error-band method.
\begin{figure*}[t!]
\includegraphics[scale=1.08]{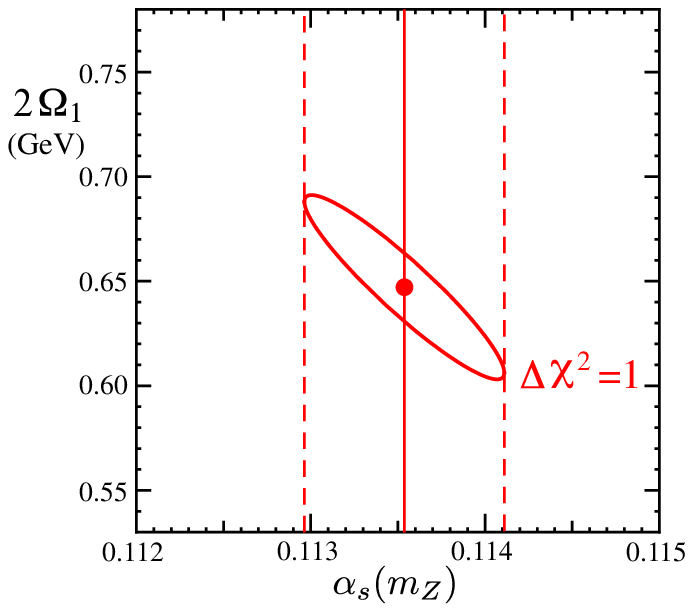}
\includegraphics[scale=0.85]{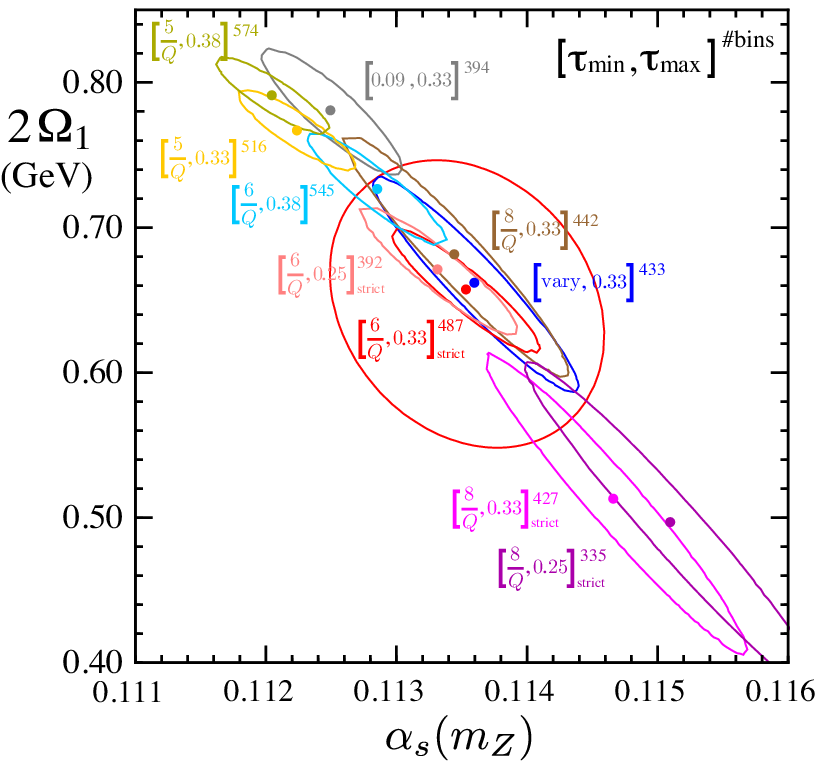}
\caption{(a) ($\chi^2_{\rm min}+1$)-ellipse for the central fit at
  N${}^3$LL${}^\prime$ order obtained from the experimental correlation matrix
  with default values for the theory scan parameters.
 (b) ($\chi^2_{\rm min}+1$)-ellipses of central fits for many different
 $\tau$-ranges in the tail region. The exponents display the number of data bins
 for each fit. The big red ellipse is the combined experimental
 and theoretical error ellipse that should be understood as 1-sigma for
 $\alpha_s(m_Z)$. 
 \label{fig:ellipsedatasets}}
\end{figure*}
It is also important to quantify the experimental uncertainties.  In
Fig.~\ref{fig:ellipsedatasets}a the ($\chi^2_{\rm min}+1$)-ellipse for the
central best fit at N${}^3$LL${}^\prime$ order is displayed. For $\alpha_s(m_Z)$
we get a purely experimental error of $(\delta\alpha_s)_{\rm exp}=0.0002$ and an
error from the variations of $\Omega_1$ of $(\delta\alpha_s)_{\rm
  \Omega_1}=0.0005$. The latter uncertainty represents the hadronization error.
Thus the theoretical uncertainties are about twice the hadronization error and
about 4 times larger than the combined statistical and (correlated) systematic
experimental errors.  The dark red ``circle'' shown in
Fig.~\ref{fig:M1alphagap}a represents the total error including experimental,
theoretical and hadronization errors. We note that to obtain stable fit results
in the $\alpha_s-\Omega_1$ plane it is essential to simultaneously fit data from
different c.m.\ energies $Q$ because there is a strong theoretical degeneracy
between $\alpha_s$ and $\Omega_1$. For each $Q$ value an increase of $\alpha_s$
can be compensated for the thrust distribution by a decrease of $\Omega_1$. The
strength of the degeneracy has, however, a strong dependence on $Q$, and can
therefore be lifted by considering data from many different $Q$ values within a
single global fit.
 
Finally, let us have a look at the dependence of the fits on the $\tau$-ranges
used for experimental data. In Fig.~\ref{fig:ellipsedatasets}b the central best
fits and the corresponding $(\chi^2_{\rm min}+1$) error ellipses for various
$\tau$-ranges are displayed. The distribution of central fits and the ellipses
is a remnant of the $\alpha_s-\Omega_1$ degeneracy just mentioned and arises
from the dependence of how the degeneracy is lifted in a global fit on the
selected $\tau$ fit range. Including more peak data at small $\tau$ leads to
smaller $\alpha_s$ (but sensitivity to the second moment $\Omega_2$ grows), and
including less data increases the experimental/hadronization errors.  The
distribution of the different best fit points represents a theoretical
uncertainty which should not be double-counted with the theory uncertainty we
already estimated from the parameter scan shown in Fig.~\ref{fig:M1alphagap}a.
This is compatible with the combined theoretical, experimental and hadronization
error from our default $\tau$-range also shown in
Fig.~\ref{fig:ellipsedatasets}b.  Our final result from our global analysis
reads
\begin{eqnarray}
\alpha_s(m_Z) &= &0.1135 \,\pm \,(0.0002)_{\rm expt}
  \,\pm\, (0.0005)_{\Omega_1} \,\pm\, (0.0009)_{\rm pert}
\nonumber\\[2mm] & 
= &0.1135 \,\pm \, (0.0011)_{\rm tot}
\,.
\end{eqnarray}

This work was supported in part by the European Community's Marie-Curie Research
Training Networks MRTN-CT-2006-035505 (HEPTOOLS), and MTRN-CT-2006-035482
(Flavianet), the Office of Nuclear Physics of the U.S.\ Department of Energy,
DE-FG02-94ER40818 and DE-FG02-06ER41449, the Alexander von Humboldt foundation, the German Academic
Exchange Service D/07/44491 and the Max-Planck-Institut f\"ur Physik guest
program.


\begin{thebibliography}{99}


\bibitem{Kluth:2006bw}
  S.~Kluth,
  Rept.\ Prog.\ Phys.\  {\bf 69}, 1771 (2006)
  [arXiv:hep-ex/0603011].

\bibitem{Farhi:1977sg}
  E.~Farhi,
  Phys.\ Rev.\ Lett.\  {\bf 39}, 1587 (1977).

\bibitem{Bauer:2002nz}
  C.~W.~Bauer, S.~Fleming and M.~E.~Luke,
  Phys.\ Rev.\  D {\bf 63}, 014006 (2000)
  [arXiv:hep-ph/0005275];
 C.~W.~Bauer, S.~Fleming, D.~Pirjol and I.~W.~Stewart,
  Phys.\ Rev.\  D {\bf 63}, 114020 (2001)
  [arXiv:hep-ph/0011336];
  C.~W.~Bauer, D.~Pirjol and I.~W.~Stewart,
  Phys.\ Rev.\  D {\bf 65}, 054022 (2002)
  [arXiv:hep-ph/0109045];
  C.~W.~Bauer et.al.,
  Phys.\ Rev.\  D {\bf 66}, 014017 (2002)
  [arXiv:hep-ph/0202088].

\bibitem{AFHMS1}
R.~Abbate, M.~Fickinger, A.H.~Hoang, V.~Mateu, I.W.~Stewart, {\it in preparation}.

\bibitem{Becher:2008cf}
  T.~Becher and M.~D.~Schwartz,
  JHEP {\bf 0807}, 034 (2008)
  [arXiv:0803.0342 [hep-ph]].

\bibitem{Catani:1992ua}
  S.~Catani, L.~Trentadue, G.~Turnock and B.~R.~Webber,
  Nucl.\ Phys.\  B {\bf 407}, 3 (1993).

\bibitem{Hoang:2008yj}
  A.~H.~Hoang, A.~Jain, I.~Scimemi and I.~W.~Stewart,
  Phys.\ Rev.\ Lett.\  {\bf 101}, 151602 (2008)
  [arXiv:0803.4214 [hep-ph]];
  [arXiv:0908.3189 [hep-ph]].

\bibitem{Hoang:2007vb}
  A.~H.~Hoang and I.~W.~Stewart,
  Phys.\ Lett.\  B {\bf 660}, 483 (2008)
  [arXiv:0709.3519 [hep-ph]].

\bibitem{Hoang:2008fs}
  A.~H.~Hoang and S.~Kluth,
  [arXiv:0806.3852 [hep-ph]].

\bibitem{Catani:1996jh}
  S.~Catani and M.~H.~Seymour,
  Phys.\ Lett.\  B {\bf 378}, 287 (1996)
  [arXiv:hep-ph/9602277].

\bibitem{GehrmannDe Ridder:2007bj}
  A.~Gehrmann-De Ridder, T.~Gehrmann, E.~W.~N.~Glover and G.~Heinrich,
  Phys.\ Rev.\ Lett.\  {\bf 99}, 132002 (2007)
  [arXiv:0707.1285 [hep-ph]].

\bibitem{Weinzierl:2009yz}
  S.~Weinzierl,
  Phys.\ Rev.\  D {\bf 80} (2009) 094018
  [arXiv:0909.5056 [hep-ph]].


\bibitem{Fleming:2007qr}
  S.~Fleming, A.~H.~Hoang, S.~Mantry and I.~W.~Stewart,
  Phys.\ Rev.\  D {\bf 77}, 074010 (2008)
  [arXiv:hep-ph/0703207];
  Phys.\ Rev.\  D {\bf 77}, 114003 (2008)
  [arXiv:0711.2079 [hep-ph]].

\bibitem{Baikov:2009bg}
  P.~A.~Baikov, K.~G.~Chetyrkin, A.~V.~Smirnov, V.~A.~Smirnov and M.~Steinhauser,
  Phys.\ Rev.\ Lett.\  {\bf 102}, 212002 (2009)
  [arXiv:0902.3519 [hep-ph]].

\bibitem{Lee:2010cg}
  R.~N.~Lee, A.~V.~Smirnov and V.~A.~Smirnov,
  JHEP {\bf 1004}, 020 (2010)
  [arXiv:1001.2887].

\end{thebibliography}
\end{document}